# Why highly expressed proteins evolve slowly

D. Allan Drummond*, Jesse D. Bloom†, Christoph Adami‡, Claus O. Wilke‡,
and Frances H. Arnold†

*Program in Computation and Neural Systems and
†Division of Chemistry and Chemical Engineering,
California Institute of Technology
Pasadena, CA 91125-4100, USA
‡Keck Graduate Institute
Claremont, CA 91711, USA

Keywords: yeast, evolutionary rate, protein misfolding, translation errors, gene duplication

Corresponding author:   D. Allan Drummond
                        Program in Computation and Neural Systems
                        California Institute of Technology
                        Mail code 210-41
                        Pasadena, CA 91125
                        Tel:  (626) 395 4664
                        Fax: (626) 568 8743
                        E-mail:  drummond@caltech.edu

Manuscript information: 30 pages, 3 figures, 1 table
Text character count: 41,244
Abstract word count: 166

# ABSTRACT

Much recent work has explored molecular and population-genetic constraints on the rate of protein sequence evolution. The best predictor of evolutionary rate is expression level, for reasons which have remained unexplained. Here, we hypothesize that selection to reduce the burden of protein misfolding will favor protein sequences with increased robustness to translational missense errors. Pressure for translational robustness increases with expression level and constrains sequence evolution. Using several sequenced yeast genomes, global expression and protein abundance data, and sets of paralogs traceable to an ancient whole-genome duplication in yeast, we rule out several confounding effects and show that expression level explains roughly half the variation in *Saccharomyces cerevisiae* protein evolutionary rates. We examine causes for expression's dominant role and find that genome-wide tests favor the translational robustness explanation over existing hypotheses that invoke constraints on function or translational efficiency. Our results suggest that proteins evolve at rates largely unrelated to their functions, and can explain why highly expressed proteins evolve slowly across the tree of life.

A central problem in molecular evolution is why proteins evolve at different rates. Protein evolutionary rates, quantified by the number of nonsynonymous nucleotide changes per site in the encoding genes, are routinely used to build phylogenetic trees, detect selection, find orthologous proteins among related species (1), and evaluate the functional importance of genes (2), yet we possess only hints of the biophysical cause of rate differences. Thirty years ago, Zuckerkandl proposed that a protein's sequence will evolve at a rate primarily determined by the proportion of its sites involved in specific functions, or its "functional density" (3). While this proposal has gained wide acceptance (2), measurement of functional density remains problematic because residues may contribute to protein function in unpredictable ways and arduous sequence-wide saturation mutagenesis and mutant characterization studies are required to ascertain these effects.

Instead, many recent studies have focused on other, more readily obtained measures which may approximate functional density. For example, protein-protein interactions presumably constrain interfacial residues, and some reports indicate that highly interactive proteins evolve slowly (4). The intuition that a protein's overall functional importance should amplify the fitness costs of mutations at sites which make subtle functional contributions has been captured in analyses of how a gene's functional category (5, 6), its essentiality for organism survival (6-8), or the fitness effect of its deletion (or "dispensability") (9, 10) correlate with evolutionary rate. In all cases, the effects under consideration explain only a small fraction (~5% or less) of the observed variation in evolutionary rate as quantified by their squared correlation coefficients $r^2$.

Surprisingly, from bacteria to mammals, the best indicator of a protein's relative evolutionary rate is the expression level of the encoding gene, measured in mRNA transcripts per cell (5, 6, 11-14). Highly expressed proteins evolve slowly, accounting for as much as 34% of rate variation in yeast (5). Moreover, after expression level is controlled for, the remaining influence of protein-protein interactions and dispensability decreases or, in some data sets, vanishes completely (15, 16). Expression level's disproportionate influence remains unexplained (5, 6, 16-19).

Significant questions have persisted about whether expression level truly determines evolutionary rate, because highly expressed proteins may possess unique structural or functional features which constrain their sequences. Paralogous gene pairs resulting from a whole-genome duplication (WGD) event, such as in the lineage of *Saccharomyces cerevisiae* (20), minimize such differences: homology ensures a similar structure, and the majority of yeast paralogs shows little, if any, difference in function (21). Analyses of evolutionary rates among paralogs have to date confirmed only a small independent role for expression level. Among a set of 185 yeast paralog pairs, evolutionary rate and expression level in mRNA molecules per cell correlated ($r^2 = 0.341$), but the correlation of rate and expression differences between members of a paralogous pair was much smaller ($r^2 = 0.046$), and no significant tendency for the higher-expressed paralog to evolve slower was found (5). A recent study which proved the whole-genome duplication in yeast (20) analyzed patterns of paralog evolutionary rates and concluded that they supported a widely cited model of evolution by gene duplication (22) in which one duplicate gene retains the ancestral function and evolves slowly, while the other evolves rapidly and acquires a new function. Such behavior would obscure the influence of other variables such as expression level on paralog evolutionary rates.

Recently, several resources have become available that allow a more thorough analysis of these issues: a set of 900 *S. cerevisiae* paralogs derived from gene synteny and traceable to the whole-genome duplication event (20), a global measurement of yeast protein abundances (23), and several additional yeast genome sequences (20, 24). Here, using this new information, we examine the strength, independence and physical basis of expression-based constraints on protein sequence evolution. We carry out a systematic analysis designed to answer several questions. How strongly does expression constrain yeast protein evolution after controlling for structure and function? What role does functional differentiation play compared to gene expression in predicting the relative evolutionary rates of duplicate genes? And, what do these correlations reveal about underlying causes of evolutionary rate differences? We introduce a novel hypothesis to explain why highly expressed proteins evolve slowly, and test this explanation against other causal hypotheses using genome-wide data. Finally, we explore whether the selective pressure we propose increases functional density, and examine the biological costs underlying it.

## MATERIALS AND METHODS

**Gene sequences**: Genome sequences for *Saccharomyces cerevisiae*, *S. kudriavzevii*, *S. paradoxus*, *S. mikatae* and *S. bayanus* were obtained from the *Saccharomyces* Genome Database (ftp://genome-ftp.stanford.edu/). The genome sequence of *Kluyveromyces waltii* was obtained from ref. (20), supplemental information.

**Identification of orthologs and paralogs**: 900 paralogous *S. cerevisiae* genes identified by synteny (20) were downloaded (20, supplemental information). Of these pairs, 290 (580 genes) were non-ribosomal proteins with a measured expression level (25) and an ortholog in *S. bayanus*, and were used in our analysis. We excluded ribosomal proteins from all analyses because they tend to be highly expressed and slow-evolving and could skew our results.

Orthologs for *S. cerevisiae* genes in members of the *Saccharomyces* genus were found by the reciprocal shortest distance (RSD) algorithm (1) with a protein-protein BLAST (26) *E*-value cutoff of $10^{-20}$, 80% minimum alignable residues, and distances computed as the number of nonsynonymous substitutions per site, dN using PAML (see below). RSD yielded 4,255 non-ribosomal *S. cerevisiae* genes with *S. bayanus* orthologs and a measured expression level, 2,790 genes with *S. mikatae* orthologs, 4,407 with *S. paradoxus* orthologs and 2,984 with *S. kudriavzevii* orthologs. The *S. paradoxus* ortholog set was expanded to include *S. cerevisiae* matches reported by Kellis *et al.* (24, supporting information).

**Expression level data**: We used gene expression data measured in mRNA molecules per cell by Holstege *et al.* (25). To estimate variability in expression level data, we used normalized fluorescence data collected using the same commercial oligonucleotide array by Cho *et al.* (27) with mean expression levels computed as described (28). Because laboratory growth media and temperatures may not reflect evolutionarily relevant environmental conditions, potentially distorting expression profiles, we repeated all analyses using each gene's codon adaptation index (CAI) (29) as an expression-level proxy (10) (see *Supporting Text* and Figure 4 in supporting information). We assume that species closely related to *S. cerevisiae* have similar expression profiles.

**Measurement of evolutionary rates**: Orthologous gene alignments were constructed from protein sequences aligned using CLUSTAL W (30). The number of nonsynonymous and synonymous substitutions per site, dN and dS, were estimated by maximum likelihood using the PAML (31) program "codeml" operating on codons.

**Statistical analysis**: We used R (32) for statistical analysis and plotting. To compute correlations on log-transformed dN data, we applied the transformation $f(k) = \log(k + 0.001)$ as in a previous study (10) to avoid excluding zeros.

## RESULTS

**Expression level and evolutionary rate**: Using genome-wide measurements of expression level (mRNA molecules per cell) and evolutionary rate (the number of nonsynonymous substitutions per site, dN) in *S. cerevisiae*, we confirm that expression level strongly predicts protein evolutionary rate. Figure 1a shows that expression level alone explains between a quarter and a third of the uncorrected variance in dN for 4,255 *S. cerevisiae* proteins with *S. bayanus* orthologs and measured expression levels (Pearson $r^2_{\text{dN-expr}} = 0.28$, $P \ll 10^{-9}$) and for the 580 paralogs (290 pairs) ($r^2_{\text{dN-expr}} = 0.31$, $P \ll 10^{-9}$). We find that the strongest simple relationship linking dN and expression is a power law (linear on a log-log scale) and that evolutionary rates span three orders of magnitude. Expression level affects evolutionary rates of duplicated and non-duplicated genes similarly.

Structural or functional differences between proteins with differing expression levels may systematically bias the dN–expression relationship. If the power-law relationship observed across paralogs holds between paralogs in a pair, the ratio of paralog expression levels should correlate linearly with the ratio of evolutionary rates on a log-log scale. Figure 1b confirms this prediction ($r^2_{dN\text{-expr}} = 0.29$, $P \ll 10^{-9}$), and demonstrates that a more limited previous analysis (5) underestimated this relationship's strength by more than six-fold.

Measurement noise attenuates correlations, possibly obscuring the strength of the relationships we have examined. For example, yeast gene expression levels measured by different groups correlate with coefficients of only 0.39 to 0.68 (28). We therefore first examined the dependence of relative inter-paralog evolutionary rate on the degree of expression level disparity, and found a dramatic association (Figure 1c). For all 290 pairs, in 192 cases the higher-expressed protein evolved slower ($P < 10^{-7}$, binomial test). Among the 19 pairs for which expression differs by at least 18-fold, all of the higher-expressed paralogs have evolved slower and $r^2_{dN\text{-expr}} = 0.67$. The dN–expression correlation can also be corrected for attenuation, allowing us to determine how much of the explainable variation in dN—variation not due to measurement noise—can be attributed to expression level. Spearman's correction for attenuation in a squared correlation coefficient is $r^2_{corr} = r^2_{dN\text{-}expr}/(r_{dN}r_{expr})$. We found that the correlation between two independent measurements of yeast gene expression using the same commercial oligonucleotide array was $r_{expr}$ = 0.72 (Pearson's $r$, 5,555 genes), and the correlation between dNs we measured using orthologs in *S. bayanus* to those measured using *S. paradoxus* orthologs was $r_{dN}$ = 0.92 (4,208 genes), yielding an overall $r^2_{corr}$ = 0.47 for the 580 paralogs and $r^2_{corr}$ = 0.42 for all 4,255 genes.

These analyses lead us to conclude that expression level accounts for roughly half of the explainable variation in yeast protein evolutionary rates, even when considering only proteins with similar structures and functions.

**Functional divergence of gene duplicates and evolutionary rate**: Are the disparate evolutionary rates in paralogous proteins a result of acquisition of new function ("neofunctionalization") in one paralog (20, 22), or do they simply reflect expression differences? Both explanations predict asymmetric paralog evolutionary rates measured against a pre-duplication relative. However, only the expression level explanation predicts that asymmetric rates will continue indefinitely, which can be measured using a post-duplication relative in which the genomic upheavals following whole-genome duplication (massive gene loss, genome rearrangements, neofunctionalization) have long since quieted.

For *S. cerevisiae*, the pre-duplication relative *K. waltii*, which diverged >100 million years ago, allows evaluation of evolutionary rates relative to a single gene descended directly from the ancestral duplicated gene (20) (Fig. 2). *S. paradoxus*, at present the closest relative of *S. cerevisiae* with a sequenced genome, with a divergence time of ~5 million years ago (24), provides a suitable post-duplication relative (Fig. 2).

We found unique *S. paradoxus* orthologs and measured expression levels for both paralogs in 73 of the 115 paralog pairs claimed to strongly support Ohno's functional divergence model (20) (as above, we excluded ribosomal proteins). In 64 of 73 cases (88%), the faster-evolving paralog relative to *K. waltii* has also evolved faster relative to *S. paradoxus*, even though roughly 100 million years have elapsed since the duplication event. (Using codon adaptation index [CAI] as a proxy for expression level, 74 of 84 pairs [88%] showed the same pattern.) In 48 of 52 pairs (92%) in which expression differs at least twofold, the higher-expressed paralog evolves slower. Finally, as Figure 1 shows, duplicated genes obey the same evolutionary rate–expression relationship as the rest of the genome, and relative expression between paralogs predicts their relative evolutionary rates.

In sum, we find little evidence that functional differentiation causes disparate evolutionary rates among duplicate genes, and plentiful evidence for the influence of expression level. A categorical consideration of neofunctionalization models is beyond our scope; we simply note that relative expression level cannot be ignored in evolutionary analyses of gene duplicates.

**Causal hypotheses**

Having established the strong and apparently independent correlation of expression level with evolutionary rate, we now turn to our central question: Why do highly expressed proteins evolve slowly? We will first attend to hypotheses offering a unified mechanistic explanation for most or all of expression level's effect, and only then address the possibility that expression level merely aggregates many independent effects to create the illusion of a single cause. In considering unified explanations, we begin by eliminating all the effects considered in the *Introduction*: previous analyses have already established that essentiality, dispensability, recombination rate, functional category, amino acid biosynthetic cost, and number or type of protein-protein interactions explain roughly 0–5% of evolutionary rate variation, while expression level accounts for more than 30%.

As Table 1 shows, the nonparametric correlation between expression and dN is twice as strong as that between expression and the rate of synonymous-site evolution (dS).  Nucleotide-level pressures such as transcription-associated mutation or DNA repair, or selection on mRNA structure or stability, cannot be the primary explanation for why highly expressed proteins evolve slowly, because they predict an equal expression-linked constraint on dS and dN.

We now consider three hypotheses for why highly expressed proteins evolve slowly.  The first, most concisely phrased by Rocha and Danchin (6), posits that each protein molecule contributes a small amount to organism fitness by performing its function, so mutations which reduce two proteins' functional output (*e.g.* catalytic rate) equally will have fitness effects weighted by the number of molecules of each protein in the cell, or their abundances, causing the more abundant protein to evolve slower.  We call this the "functional loss" hypothesis.  Note that a highly expressed protein (whose encoding gene is transcribed at high levels) can have a low abundance (if the mRNA is translated infrequently or the protein is rapidly turned over), and vice versa.  The second hypothesis, due to Akashi (17, 18), holds that because increased expression level leads to selection for synonymous codons that are translated faster or more accurately, nonsynonymous mutations to translationally less efficient codons may be evolutionarily disfavored, slowing the rate of amino acid sequence change.  We call this the "translational efficiency" hypothesis.

We advance a third hypothesis based on a simple observation: to reduce the number of proteins which misfold due to translation errors, selection can act both on the nucleotide sequence, to increase translational accuracy by optimizing codon usage (33), and on the amino acid sequence, to increase the number of proteins which fold properly despite mistranslation (Fig. 3). We call this increased tolerance for translational missense errors "translational robustness." At the canonical ribosomal error rate of five errors per 10,000 codons translated (34), approximately 19% of average-length yeast proteins (415 amino acids) contain a missense error, and these errors may cause misfolding (35). Proteins vary in their tolerance for amino acid substitutions (36), providing the necessary raw material for evolution, while misfolded-protein aggregation and toxicity (35, 37) and production of non-functional protein (38) impose burdens on most cellular metabolisms, providing selective pressure. So long as translationally robust sequences are comparatively rare, intensified selection pressure resulting from increased expression level will slow the rate of amino acid substitution in higher-expressed proteins.

These three hypotheses differ in important ways. The functional loss hypothesis points to loss of protein function as the key cost constraining evolution. The translational efficiency hypothesis states that the protein sequence is constrained as a side effect of selection on the mRNA sequence. And the translational robustness hypothesis instead implicates the direct costs of misfolded proteins, independent of function. These hypotheses make testable and opposing predictions, which we now consider.

**Functional loss versus translational robustness:** Given two proteins with differing abundances $A > a$, measured in protein molecules per cell, but oppositely differing expression levels $x < X$, measured in mRNA molecules per cell, the functional loss hypothesis predicts $dN_{Ax} < dN_{aX}$: the more abundant protein will evolve slower. By contrast, the translational robustness hypothesis states that fitness costs are dominated by translation-error-induced misfolding, leading to the opposite prediction ($dN_{Ax} > dN_{aX}$), because despite $Ax$'s higher abundance, $aX$'s higher expression level suggests more frequent translation and turnover (39).

We tested these competing predictions using a recent global analysis of protein abundance in yeast (23). Ten thousand unique pairs of yeast proteins for which one member had a higher expression level and a lower abundance than the other were assembled at random. In 5,579 of 10,000 pairs, the more abundant but lower-expressed protein evolved faster ($dN_{Ax} > dN_{aX}$, $P \ll 10^{-9}$, binomial test) consistent with translational robustness but contradicting the functional loss hypothesis. When we sampled pairs with at least a twofold difference in each measure, limiting the influence of measurement noise, 5,430 of 10,000 pairs showed the same pattern ($P \ll 10^{-9}$). Among synteny-derived paralog pairs, 25 of 48 showed the same pattern (not significant), as did 7 of 8 pairs with twofold differences ($P < 0.05$). Using CAI as an expression proxy (see *Methods*), 6,262 of 10,000 pairs ($P \ll 10^{-9}$) and 17 of 20 paralog pairs ($P < 0.002$) also showed the same pattern. These results suggest that the number of translation events, a correlate of expression level and CAI, is a better predictor of relative protein evolutionary rates than the number of functional protein molecules.

The functional loss hypothesis rests on the supposition that protein molecules contribute roughly the same amount to organism fitness through their biological function, so that less-abundant proteins are less important to organism fitness. We find this assumption difficult to accept on biochemical grounds. Protein abundance seems to depend mainly on substrate or target availability, which has no obvious relationship to fitness contribution. For example, most gene regulatory proteins and DNA polymerases have only a few hundred targets and correspondingly low cellular abundances, yet play crucial cellular roles. While cells seem unlikely to invest in synthesis of high-abundance proteins without a comparably high return, the inference that low-expression proteins generate low fitness returns does not follow. Accordingly, under the functional loss hypothesis, we should expect low-expression proteins to span the range of evolutionary rates, while high-expression proteins evolve under a more uniformly tight constraint. Instead, in yeast, the slowest-evolving low-expression proteins evolve an order of magnitude more rapidly than their highly expressed counterparts (Fig. 1a). This pattern again supports translational robustness, which supposes that, while folded proteins may confer widely varying fitness benefits, misfolded polypeptides impose similar costs.

**Translational efficiency versus translational robustness**: Pressure to retain translationally efficient preferred codons will constrain synonymous evolution (dS) and, as a consequence, protein evolution (dN). Pressure for translationally efficient amino acids (18) would bias amino acid preferences at aligned positions in high- and low-expression paralogs. By contrast, translational robustness predicts that the dS and dN constraints reflect two independent points of selection (Figure 3) and that no consistent translational preference for either codons or amino acids is required to explain the dN trend.

To assess the protein-level constraint attributable to selection for preferred codons, which is strongest at functionally important and conserved sites (33), we computed evolutionary rates using the portions of genes consisting only of unpreferred codons. Because those sites most constrained by codon preference are removed in these reduced genes, the codon preference hypothesis predicts that the correlation of expression level with dS and dN should vanish. Translational robustness hypothesizes a direct constraint on the amino acid sequence, so the dN–expression correlation should remain strong while the dS–expression correlation vanishes, essentially an impossibility if synonymous-site selection for translational efficiency governs protein evolution. Using sets of aligned *S. cerevisiae*–ortholog genes (see *Methods*), we discarded all aligned codons except those where the "relative adaptedness" (29) of the *S. cerevisiae* codon was less than 0.5. We then recomputed dN, dS and their expression correlations using these reduced genes, discarding genes with fewer than 30 codons or dS values of 3.0 or larger.

Table 1 shows that after removal of preferred codons, the reduced genes showed only slightly reduced dN–expression correlations, while the dS–expression correlations all became insignificant or, in the case of *S. paradoxus*, reversed direction. We found similar results using CAI as an expression proxy (see Table 2, supporting information). These results demonstrate that expression-linked synonymous selection is concentrated at sites bearing preferred codons and that sites showing no such selection still show strong protein-level constraint, consistent with selection for translational robustness.

Translational efficiency selection on amino acids predicts asymmetric substitution of one amino acid for another in highly expressed proteins. If two amino acids $x$ and $y$ have efficiencies $x < y$, then at aligned positions in paralogs where both $x$ and $y$ occur, $y$ should disproportionately appear in the higher-expressed paralog. We tabulated these pairwise frequencies in the 580 paralogs analyzed in Figure 1 and assessed statistical significance using a binomial test with the false-discovery-rate correction for multiple tests (40). All residue pairs appeared in our data set, but no pairs showed asymmetries at the 1% or 5% levels.

As a control, we performed the same test using synonymous codons, and found that 21 codon pairs showed significant asymmetries at the 1% level, invariably favoring the codon with higher relative adaptedness in the higher-expressed paralog (Table 3, supporting information). Of the 21 favored codons, 17 were unique and encoded 13 of the 18 amino acids with synonymous codons.

Our results offer no support for translational efficiency selection on amino acids, but confirm such selection on synonymous codons, though with little consequence for dN. Although translational efficiency selection may constrain amino acid sequences to some degree, it cannot explain why highly expressed yeast proteins evolve slowly.

**Expression level is a master causal variable**: We now consider the possibility that many variables (e.g., dispensability, number of protein-protein interactions, amino acid biosynthetic cost, codon preference, recombination rate) independently exert small but cumulatively severe constraining effects on protein sequence evolution, and expression level's influence derives from its relationships to each of these variables. While such a possibility cannot be ruled out, several observations make it unlikely.

First, expression level is a major determinant of most of the candidate variables: high expression causes decreased dispensability (41), causes more experimentally detected interactions (15), increases pressure for cheaper proteins and higher translational efficiency (17), and, through increased transcription, causes exposed chromatin structures that are hotspots for recombination. No reverse mechanisms have been proposed by which these variables cause genes to become highly expressed.

Second, as we have noted earlier, the degree to which these variables appear to influence evolutionary rate becomes small or even disappears after controlling for expression level. This trend holds for protein-protein interactions (4, 15), recombination rates (42), and amino acid cost in bacteria (6), as well as essentiality, dispensability, network centrality, and gene length (43).

## DISCUSSION

We have provided evidence that expression level is the dominant determinant of evolutionary rate in *S. cerevisiae* genes. Our results show that *i)* expression level explains roughly half the variation in gene evolutionary rates; *ii)* expression level affects evolutionary rates of duplicated and singleton genes similarly; *iii*) once variability in expression level is accounted for, the higher-expressed member of a paralog pair is disproportionately likely to evolve slower; *iv*) asymmetric evolutionary rates in duplicated genes persist over tens of millions of years, consistent with expression-level differences but not neofunctionalization; and *v*) expression level appears to influence evolutionary rate through the number of translation events rather than cellular protein abundance, constraining the protein sequence directly rather than through translational efficiency selection.

We have introduced a general hypothesis to explain why highly expressed proteins evolve slowly: selection against the expression-level-dependent cost of misfolded proteins favors rare protein sequences which fold properly despite translation errors (Fig. 3). Tests comparing the opposing predictions of this translational robustness hypothesis to two previously advanced alternative hypotheses show that genome-wide yeast data support the predictions of translational robustness and contradict the alternatives. Our hypothesis contradicts the intuitive notion that highly expressed proteins evolve slowly because they are more functionally important, perhaps explaining why more direct measures of functional importance, such as essentiality and dispensability, explain far less variation in evolutionary rates. The hypothesis also provides an explanation for the widely observed correlation between dN and dS (19): Figure 3 indicates how one cost (misfolding) can be counteracted in two ways (translational accuracy, slowing dS, and translational robustness, slowing dN).

Would more translationally robust proteins have a higher functional density (3)?  Consider URA5 and URA10 (orotate phosphoribosyltransferases 1 and 2), paralogs with similar functions which differ 60+-fold in expression and 6-fold in evolutionary rate.  Do we expect URA5 to have a larger proportion of its residues involved in specific functions?  The translational robustness hypothesis suggests not.  Instead, functionally unconstrained residues may be more carefully selected to preserve the protein's native structure after missense substitutions in URA5 than in URA10.  These residues would contribute to fitness not by aiding in URA5's function, but by preventing the burdensome misfolding of mistranslated polypeptides.  Thus the fitness density of a protein, the proportion of residues under meaningful natural selection, can be larger than the functional density, and directly determines the rate of sequence evolution.

Functional constraints slow evolution at certain sites; our results suggest that these constraints operate on a sequence-wide background rate determined largely by expression.  Expression patterns as well as levels may impose additional constraints if highly expressed proteins have unique cellular localization or cell-cycle expression profiles.

How large are the costs underlying translational robustness? We can make a crude general estimate. As mentioned above, roughly 19% of average-length yeast proteins will contain a missense error at typical ribosomal error rates. For diverse proteins, 20–65% of amino acid substitutions lead to inactivation (36, 44), generally due to misfolding (36). Consequently, 4–12% of a typical protein species would be expected to misfold due to missense errors. Because yeast protein abundances span five orders of magnitude (23), the fitness impact of error-induced misfolding could range widely. If we assume a 5% misfolding rate, the number of misfolded protein molecules ranges from negligible, as for the ~3 misfolded molecules to generate the measured cellular complement of 64 molecules of DSE4 (endo-1,3-β-glucanase), to potentially devastating, as for the ~63,000 misfolded molecules required to generate 1.26 million molecules per cell of the H$^+$-transporting P-type ATPase PMA1 (23). The latter misfolded species would be more abundant than 97% of yeast proteins (23). We have neglected protein turnover, a further cost multiplier. (We have also neglected the misfolding of error-free proteins; a likely biophysical mechanism for increasing translational robustness will also mitigate stochastic misfolding [see below].) Protein misfolding generates highly toxic species capable of killing cells in a concentration-dependent manner (45), so increased translational robustness in highly expressed proteins may reflect pressure for survival as well as efficiency.

Can selection for accuracy through codon preference eliminate (or make negligible) such error-induced misfolding costs? While codon preference cannot counter mistranslation due to misacylation of tRNAs and transcription errors, both of which occur at frequencies approaching those of missense errors (34), experimental measurements of a 4- to 9-fold reduction in missense errors from preferred codons have been reported (46). Assuming all preferred codons are translated 10-fold more accurately than nonpreferred codons, how much accuracy improvement can we expect? Randomly selecting codons produces genes containing ~35% preferred codons, while the most highly expressed genes have >80% preferred codons (only 9 of the 4,255 yeast genes we analyzed contain >90% preferred codons). Even if translational error-rate measurements reflect the worst case of codon-randomized genes, the maximum accuracy gain in the most optimized genes is roughly five-fold. In the case of PMA1 (86% preferred codons), such a reduction would still leave thousands of misfolded proteins from this single gene to burden the cell. While that level of misfolding may represent the "cost of doing business" for the cell, such an argument assumes that mutant versions of PMA1 carried by evolutionary competitors tolerate equivalent numbers of translation errors and generate similar costs. Because a protein's tolerance to substitutions can in some cases be significantly altered with a single mutation (36), we suspect this assumption is rarely justified. Given variability in misfolding, natural selection will then favor those mutants whose costs undercut their competitors'.

A counterintuitive prediction of the translational robustness hypothesis is that selection for proteins that are more tolerant to amino acid change yields underlying genes that appear less tolerant to nucleotide change (because they evolve slowly). How is this result possible? Consider a hypothetical allele of PMA1 for which only 0.1% (~1,000 molecules) of translated proteins misfold due to errors. A nonsynonymous genetic mutation yielding a functionally equivalent mutant protein that misfolds 5% of the time, producing ~50,000 potentially toxic proteins, would be evolutionarily disfavored relative to the wildtype due to increased misfolding costs without showing any functional difference. Thus the wildtype, despite encoding a highly robust protein which retains function after most mutations, will appear mutationally fragile over evolutionary time. A striking example of this robust-molecule/fragile-gene behavior may be found in ribulose-1,5-bisphosphate carboxylase/oxygenase (Rubisco), perhaps the most abundant protein on Earth and a rigidly conserved, generally essential enzyme for which genetic studies have nonetheless been hampered by the difficulty of finding inactivating missense mutations (47).

How might translational robustness manifest itself biophysically? We can offer only a speculation. Because most substitutions destabilize the native structure of a protein, modest increases in thermodynamic stability broaden the spectrum of substitutions a protein can tolerate before misfolding (36), increasing fitness so long as function is not compromised. Pressure for increased stability in highly expressed proteins would restrict the set of evolutionarily viable sequences and slow sequence evolution as a consequence.

## ACKNOWLEDGMENTS

This work was supported by NIH National Research Service Award 5 T32 MH19138 (to D.A.D.) and an HHMI predoctoral fellowship (to J.D.B.).

## REFERENCES

1. Wall, D. P., Fraser, H. B. & Hirsh, A. E. (2003) *Bioinformatics* **19,** 1710-1711.
2. Graur, D. & Li, W.-H. (2000) *Fundamentals of Molecular Evolution* (Sinauer Associates, Inc., Sunderland, MA).
3. Zuckerkandl, E. (1976) *J Mol Evol* **7,** 167-183.
4. Fraser, H. B., Hirsh, A. E., Steinmetz, L. M., Scharfe, C. & Feldman, M. W. (2002) *Science* **296,** 750-2.
5. Pál, C., Papp, B. & Hurst, L. D. (2001) *Genetics* **158,** 927-31.
6. Rocha, E. P. & Danchin, A. (2004) *Mol Biol Evol* **21,** 108-16.
7. Hurst, L. D. & Smith, N. G. (1999) *Curr Biol* **9,** 747-50.
8. Jordan, I. K., Rogozin, I. B., Wolf, Y. I. & Koonin, E. V. (2002) *Genome Res* **12,** 962-8.
9. Hirsh, A. E. & Fraser, H. B. (2001) *Nature* **411,** 1046-9.

10. Wall, D. P., Hirsh, A. E., Fraser, H. B., Kumm, J., Giaever, G., Eisen, M. B. & Feldman, M. W. (2005) *Proc Natl Acad Sci U S A* **102,** 5483-8.
11. Herbeck, J. T., Wall, D. P. & Wernegreen, J. J. (2003) *Microbiology* **149,** 2585-96.
12. Sharp, P. M. (1991) *J Mol Evol* **33,** 23-33.
13. Duret, L. & Mouchiroud, D. (2000) *Mol Biol Evol* **17,** 68-74.
14. Subramanian, S. & Kumar, S. (2004) *Genetics* **168,** 373-81.
15. Bloom, J. D. & Adami, C. (2003) *BMC Evol Biol* **3,** 21.
16. Pál, C., Papp, B. & Hurst, L. D. (2003) *Nature* **421,** 496-7; discussion 497-8.
17. Akashi, H. (2001) *Curr Opin Genet Dev* **11,** 660-6.
18. Akashi, H. (2003) *Genetics* **164,** 1291-303.
19. Marais, G., Domazet-Loso, T., Tautz, D. & Charlesworth, B. (2004) *J Mol Evol* **59,** 771-9.
20. Kellis, M., Birren, B. W. & Lander, E. S. (2004) *Nature* **428,** 617-24.
21. Seoighe, C. & Wolfe, K. H. (1999) *Curr Opin Microbiol* **2,** 548-54.
22. Ohno, S. (1970) *Evolution by Gene Duplication* (Allen and Unwin, London).
23. Ghaemmaghami, S., Huh, W. K., Bower, K., Howson, R. W., Belle, A., Dephoure, N., O'Shea, E. K. & Weissman, J. S. (2003) *Nature* **425,** 737-41.
24. Kellis, M., Patterson, N., Endrizzi, M., Birren, B. & Lander, E. S. (2003) *Nature* **423,** 241-54.
25. Holstege, F. C., Jennings, E. G., Wyrick, J. J., Lee, T. I., Hengartner, C. J., Green, M. R., Golub, T. R., Lander, E. S. & Young, R. A. (1998) *Cell* **95,** 717-28.
26. Altschul, S., Madden, T., Schaffer, A., Zhang, J. H., Zhang, Z., Miller, W. & Lipman, D. (1998) *The FASEB journal* **12,** A1326.
27. Cho, R. J., Campbell, M. J., Winzeler, E. A., Steinmetz, L., Conway, A., Wodicka, L., Wolfsberg, T. G., Gabrielian, A. E., Landsman, D., Lockhart, D. J. & Davis, R. W. (1998) *Mol Cell* **2,** 65-73.
28. Coghlan, A. & Wolfe, K. H. (2000) *Yeast* **16,** 1131-45.
29. Sharp, P. M. & Li, W. H. (1987) *Nucleic Acids Res* **15,** 1281-95.
30. Thompson, J. D., Higgins, D. G. & Gibson, T. J. (1994) *Nucleic Acids Res* **22,** 4673-80.
31. Yang, Z. H. (1997) *Computer Applications in the Biosciences* **13,** 555-556.
32. Ihaka, R. & Gentleman, R. (1996) *Journal of Computational and Graphical Statistics* **5,** 299-314.
33. Akashi, H. (1994) *Genetics* **136,** 927-35.
34. Parker, J. (1989) *Microbiol Rev* **53,** 273-98.
35. Goldberg, A. L. (2003) *Nature* **426,** 895-9.
36. Bloom, J. D., Silberg, J. J., Wilke, C. O., Drummond, D. A., Adami, C. & Arnold, F. H. (2005) *Proc Natl Acad Sci U S A* **102,** 606-611.

37. Ellis, R. J. & Pinheiro, T. J. (2002) *Nature* **416,** 483-4.
38. Dong, H., Nilsson, L. & Kurland, C. G. (1995) *J Bacteriol* **177,** 1497-504.
39. Greenbaum, D., Colangelo, C., Williams, K. & Gerstein, M. (2003) *Genome Biol* **4,** 117.
40. Benjamini, Y. & Hochberg, Y. (1995) *Journal of the Royal Statistical Society Series B* **57,** 289-300.
41. Gu, Z., Steinmetz, L. M., Gu, X., Scharfe, C., Davis, R. W. & Li, W. H. (2003) *Nature* **421,** 63-6.
42. Pál, C., Papp, B. & Hurst, L. D. (2001) *Mol Biol Evol* **18,** 2323-6.
43. Drummond, D. A., Raval, A. & Wilke, C. O. (2005) *arXiv: q-bio.PE/0506011*.
44. Guo, H. H., Choe, J. & Loeb, L. A. (2004) *Proc Natl Acad Sci U S A* **101,** 9205-10.
45. Bucciantini, M., Giannoni, E., Chiti, F., Baroni, F., Formigli, L., Zurdo, J., Taddei, N., Ramponi, G., Dobson, C. M. & Stefani, M. (2002) *Nature* **416,** 507-11.
46. Precup, J. & Parker, J. (1987) *J Biol Chem* **262,** 11351-5.
47. Spreitzer, R. J. (1993) *Annual Review of Plant Physiology and Plant Molecular Biology* **44,** 411-434.
48. Rokas, A., Williams, B. L., King, N. & Carroll, S. B. (2003) *Nature* **425,** 798-804.
49. Kurtzman, C. P. & Robnett, C. J. (2003) *FEMS Yeast Research* **3,** 417-432.

**FIGURE LEGENDS**

**Figure 1**. Expression level governs gene and paralog evolutionary rates in *S. cerevisiae*. **a**, Highly expressed proteins evolve more slowly, and paralogs mirror the genome-wide pattern. Evolutionary rates measured relative to *S. bayanus* for 4,255 *S. cerevisiae* genes (□) and 580 paralogous genes (■) correlate with expression levels. Lines show best log-log linear fit. For all genes (dotted line), $r^2$ = 0.28, $P \ll 10^{-9}$; for paralogs (solid line), $r^2 = 0.31$, $P \ll 10^{-9}$. **b**, Within a paralog pair, the ratio of expression levels correlates with the ratio of evolutionary rates ($r^2 = 0.29$, $P \ll 10^{-9}$), as predicted from the log-log linear relationship in **a**. Each pair generates two ratio points, making the plot symmetrical. **c**, Relative expression level determines relative evolutionary rate. The percentage of pairs in which the higher-expressed paralog evolves slower are shown as a function of minimum paralog pair expression ratio (■). Point areas are proportional to the number of included pairs.

**Figure 2**. Phylogenetic relationships between analyzed yeast species. Relationships follow ref. (48), branch lengths indicate nucleotide substitution distances from ref. (49), and the indicated time of the whole-genome duplication follows ref. (20).

**Figure 3**. Translational selection against the cost of misfolded proteins can act at two distinct points. Messenger RNA (left) may be translated without errors to produce a folded protein (top); if an error is made, the resulting protein may still fold properly, or may misfold and undergo degradation (right). Selection can act at **A** to increase the proportion of error-free proteins through codon preference (translational accuracy), and also at **B** to increase the proportion of proteins that fold despite errors (translational robustness). We neglect misfolding of error-free proteins (see text).

**Table 1**: Evolutionary rate vs. expression correlations (Kendall's τ) relative to four yeast species for *S. cerevisiae* genes, including and excluding preferred codons.

| Ortholog (# of genes) | All codons | | Codons with relative adaptedness < 0.5 | |
|---|---|---|---|---|
| | dN–expr.† | dS–expr. | dN–expr. | dS–expr. |
| *S. bayanus* (2,614) | τ = −0.300*** | τ = −0.181*** | τ = −0.273*** | τ = −0.010 |
| *S. mikatae* (2,102) | τ = −0.335*** | τ = −0.163*** | τ = −0.302*** | τ = −0.009 |
| *S. paradoxus* (4,383) | τ = −0.340*** | τ = −0.153*** | τ = −0.303*** | τ = +0.046** |
| *S. kudriavzevii* (2,193) | τ = −0.340*** | τ = −0.162*** | τ = −0.314*** | τ = −0.004 |

†Significance codes: *, $P < 10^{-2}$; **, $P < 10^{-4}$; ***, $P < 10^{-6}$.

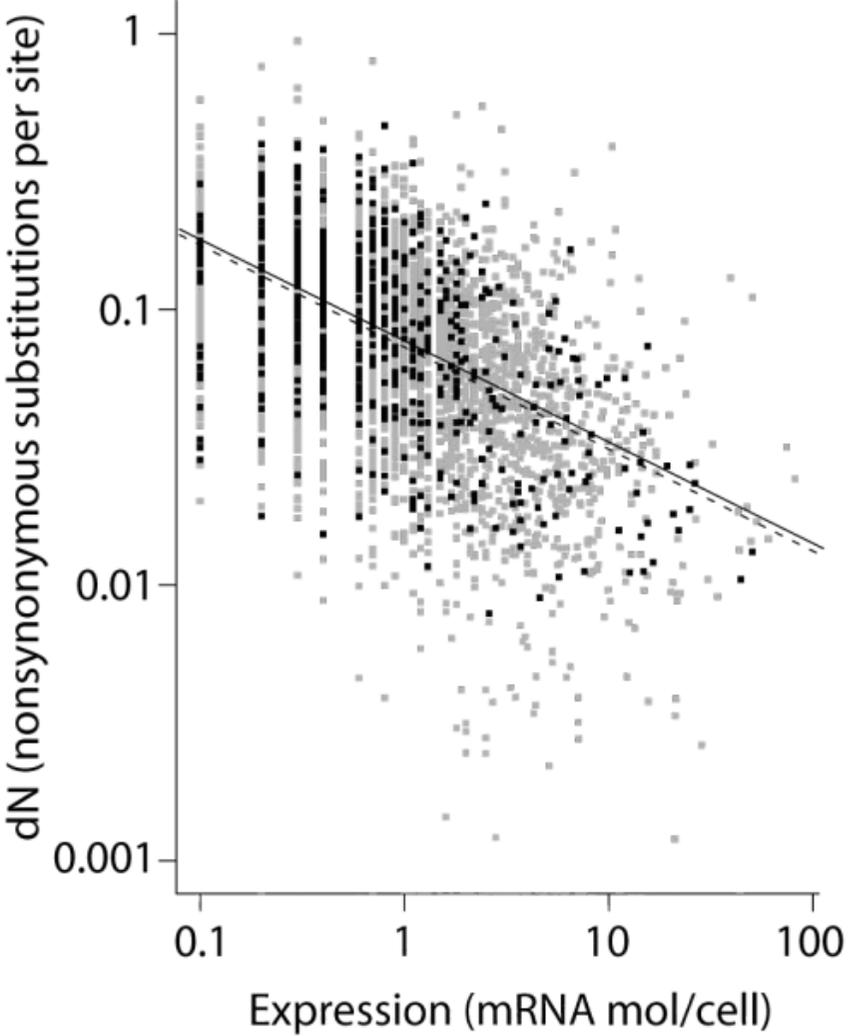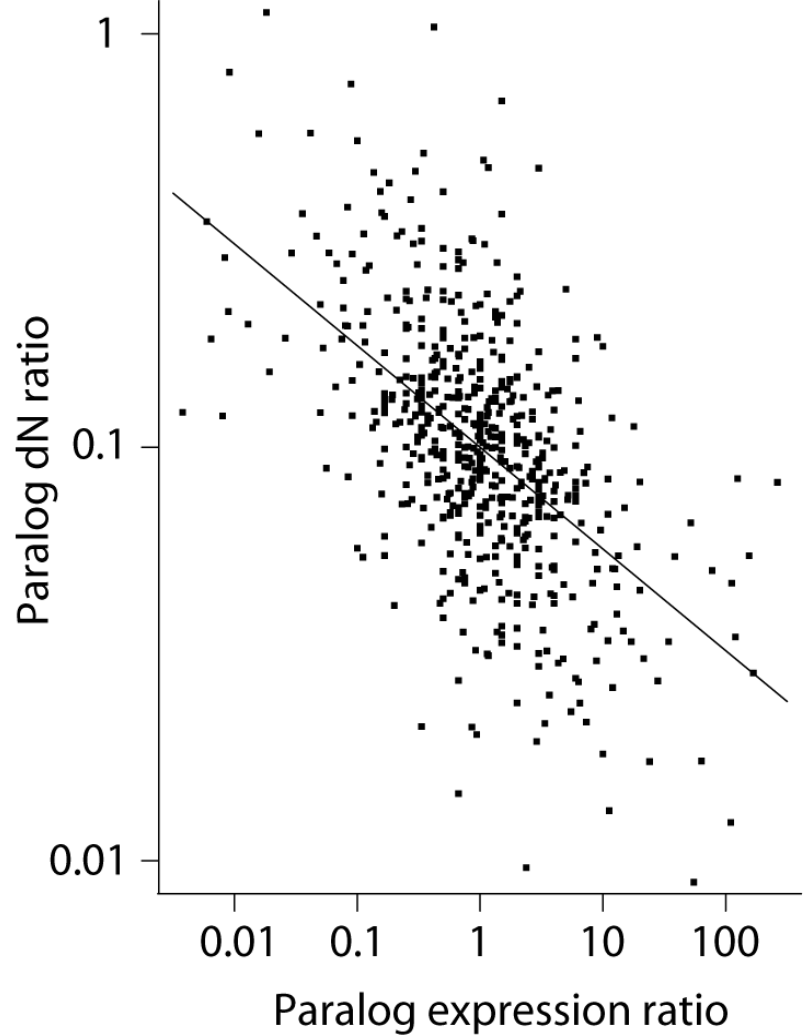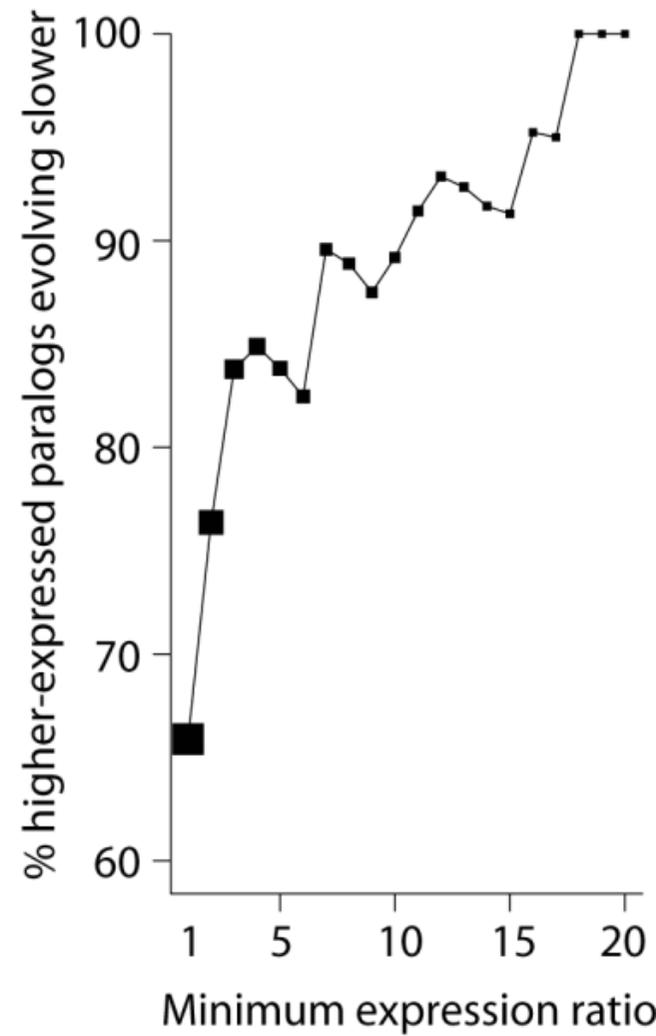

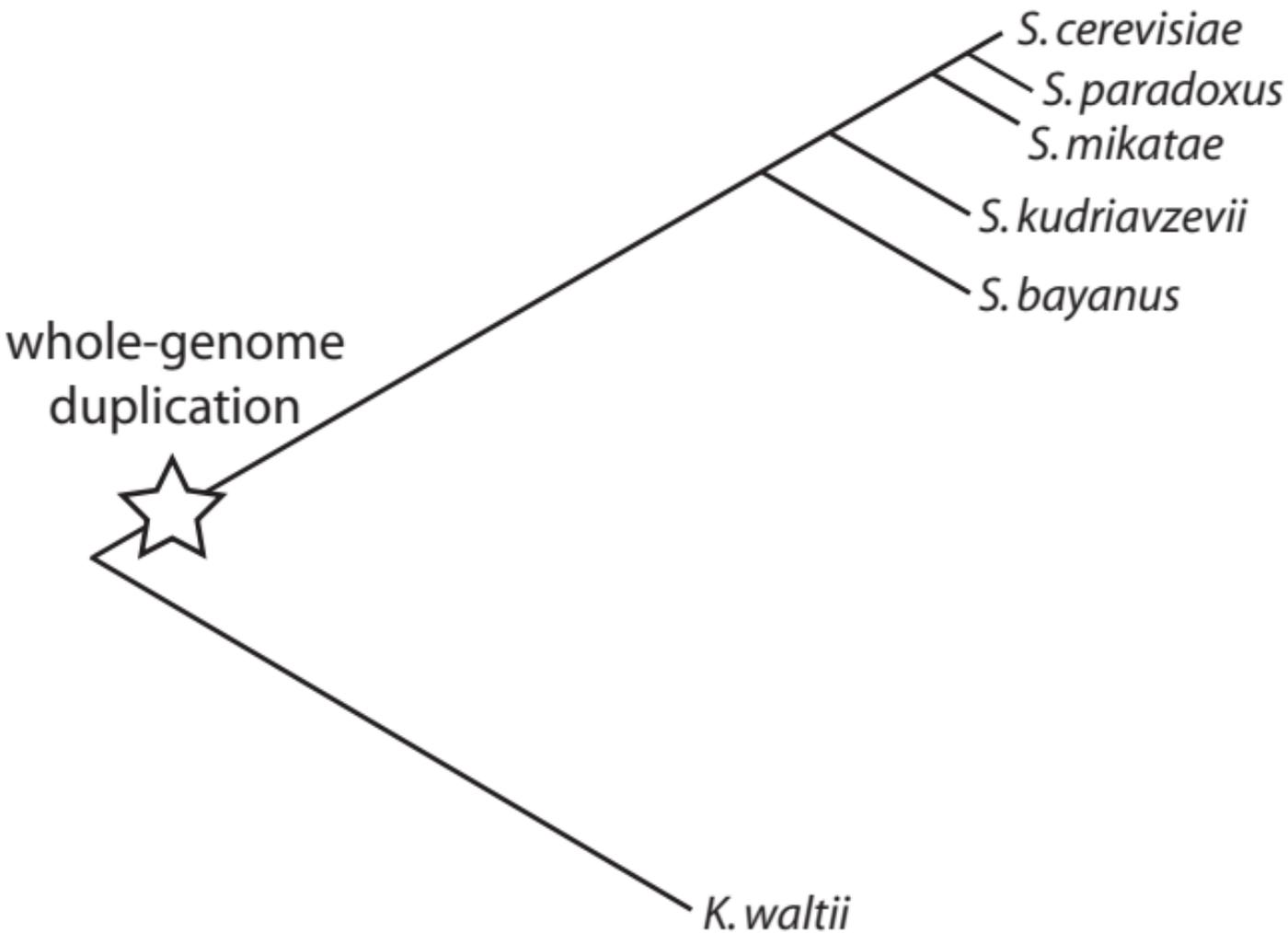

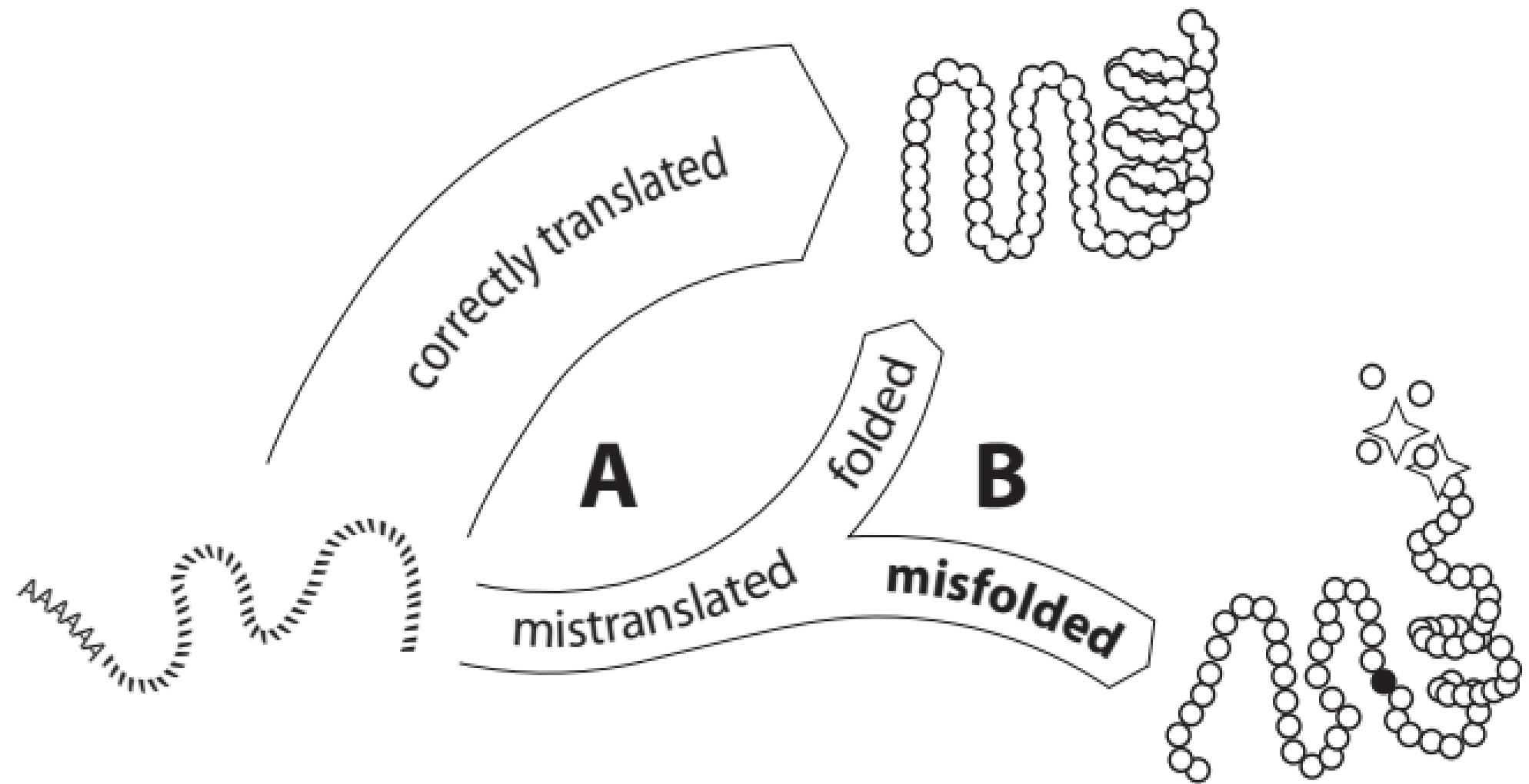

**Supporting Information**

**Figure Legend**

**Figure 4**. Expression level as estimated by the codon adaptation index (CAI) reveals evolutionary rate relationships similar to those found using expression level. **a**, Highly expressed proteins evolve more slowly, and paralogs mirror the genome-wide pattern. Evolutionary rates measured relative to *S. bayanus* for 4,534 *S. cerevisiae* genes (■) and 650 paralogous genes (■) correlate with CAI. Lines show best log-log linear fit. For all genes (dotted line), $r^2 = 0.27$, $P \ll 10^{-9}$; for paralogs (solid line), $r^2 = 0.38$, $P \ll 10^{-9}$. **b**, Within a paralog pair, the ratio of expression levels correlates with the ratio of evolutionary rates ($r^2 = 0.31$, $P \ll 10^{-9}$), as predicted from the log-log linear relationship in **a**. Each pair generates two ratio points, making the plot symmetrical. **c**, Relative CAI governs relative evolutionary rate. The percentage of pairs in which the higher-expressed paralog evolves slower are shown as a function of minimum paralog pair CAI ratio (■). Point areas are proportional to the number of included pairs.

**Table 2**: Evolutionary rate vs. CAI correlations (Kendall's τ) relative to four yeast species for *S. cerevisiae* genes, including and excluding preferred codons.

| Ortholog (# of genes) | All codons | | Codons with relative adaptedness < 0.5 | |
|---|---|---|---|---|
| | dN–CAI[†] | dS–CAI | dN–CAI | dS–CAI |
| *S. bayanus* (2,613) | τ = −0.268*** | τ = −0.096*** | τ = −0.233*** | τ = +0.099*** |
| *S. mikatae* (2,108) | τ = −0.321*** | τ = −0.050* | τ = −0.281*** | τ = +0.107*** |
| *S. paradoxus* (4,656) | τ = −0.326*** | τ = −0.068*** | τ = −0.277*** | τ = +0.146*** |
| *S. kudriavzevii* (2,340) | τ = −0.281*** | τ = −0.050* | τ = −0.245*** | τ = +0.112*** |

[†]Significance codes: *, $P < 10^{-2}$; **, $P < 10^{-4}$; ***, $P < 10^{-6}$.

**Table 3:** Significant asymmetries in synonymous codon usage between high- and low-expressed paralogs at aligned positions reflects relative adaptedness.

| Codon | | Amino acid | #(low, high) | | $P^\dagger$ | Rel. adaptedness | |
|---|---|---|---|---|---|---|---|
| x | y | | #(x,y) | #(y,x) | | x | y |
| GCA | GCC | A | 281 | 215 | * | 0.015 | 0.316 |
| GCA | GCT | A | 479 | 312 | *** | 0.015 | 1.000 |
| GAG | GAA | E | 1081 | 869 | ** | 0.016 | 1.000 |
| GGC | GGT | G | 463 | 350 | * | 0.020 | 1.000 |
| GGG | GGT | G | 306 | 187 | ** | 0.004 | 1.000 |
| GGA | GGT | G | 552 | 346 | *** | 0.002 | 1.000 |
| CAT | CAC | H | 391 | 294 | * | 0.245 | 1.000 |
| ATA | ATC | I | 364 | 266 | * | 0.003 | 1.000 |
| AAA | AAG | K | 1315 | 1130 | * | 0.135 | 1.000 |
| CTT | TTA | L | 314 | 241 | * | 0.006 | 0.117 |
| TTA | TTG | L | 855 | 730 | * | 0.117 | 1.000 |
| AAT | AAC | N | 972 | 830 | * | 0.053 | 1.000 |
| CCT | CCA | P | 554 | 433 | * | 0.047 | 1.000 |
| CCG | CCA | P | 254 | 172 | * | 0.002 | 1.000 |
| AGG | CGT | R | 74 | 42 | * | 0.003 | 0.137 |
| AGG | AGA | R | 511 | 377 | ** | 0.003 | 1.000 |
| ACA | ACT | T | 413 | 315 | * | 0.012 | 0.921 |
| GTA | GTT | V | 309 | 236 | * | 0.002 | 1.000 |
| GTA | GTC | V | 171 | 117 | * | 0.002 | 0.831 |
| GTG | GTT | V | 322 | 212 | ** | 0.018 | 1.000 |
| TAT | TAC | Y | 886 | 692 | ** | 0.071 | 1.000 |

$^\dagger$Binomial probability with false-discovery-rate correction for multiple tests. Significance codes: *, $P < 10^{-2}$; **, $P < 10^{-4}$; ***, $P < 10^{-6}$.

**Supporting Text**

We repeated each of our analyses using a gene's codon adaptation index (CAI) as a proxy for its expression level, allowing us to expand the coverage of our tests and to eliminate the dependence of expression measurements on particular growth conditions. Figure 4 shows that all the trends we identified in Figure 1 remain highly significant using the CAI proxy. For the 325 pairs with *S. bayanus* orthologs, 224 of the higher-CAI paralogs evolved slower than their lower-CAI counterpart ($P \ll 10^{-9}$, binomial test). Table 2 demonstrates that elimination of preferred codons obliterates the negative correlation between CAI and dS across multiple species.

As discussed in the main text, we examined asymmetries between the frequencies of synonymous codons ($x$ and $y$) at aligned positions in two paralogs. We counted the number of times $x$ appeared in the lower-expressed paralog while $y$ appeared at the aligned position in the higher-expressed paralog, #($x$, $y$), and the number of times $y$ appeared in the lower-expressed paralog while $x$ appeared at the aligned position in the higher-expressed paralog, #($y$, $x$). Deviations from chance were assessed by the binomial test with the false-discovery-rate correction for multiple tests (1). Codons favored at the 1% level are reported in Table 3. In all cases, the codon with higher relative adaptedness (2) is favored in higher-expressed paralogs. At the 5% level, 38 significant pairs were found, corresponding to 25 unique favored codons which encode 15 of 18 amino acids with synonymous codons. At either significance level, no amino acid pairs showed asymmetries when subjected to the same test.

## REFERENCES

1. Benjamini, Y. & Hochberg, Y. (1995) *Journal of the Royal Statistical Society Series B* **57,** 289-300.
2. Sharp, P. M. & Li, W. H. (1987) *Nucleic Acids Res* **15,** 1281-95.

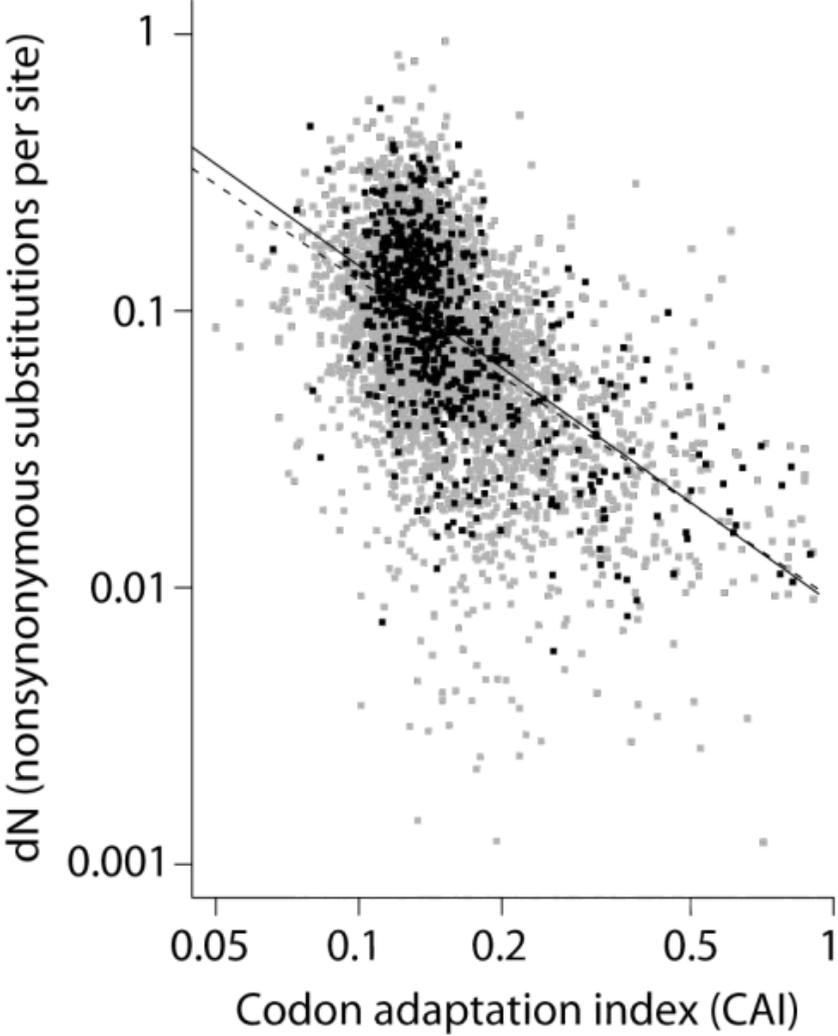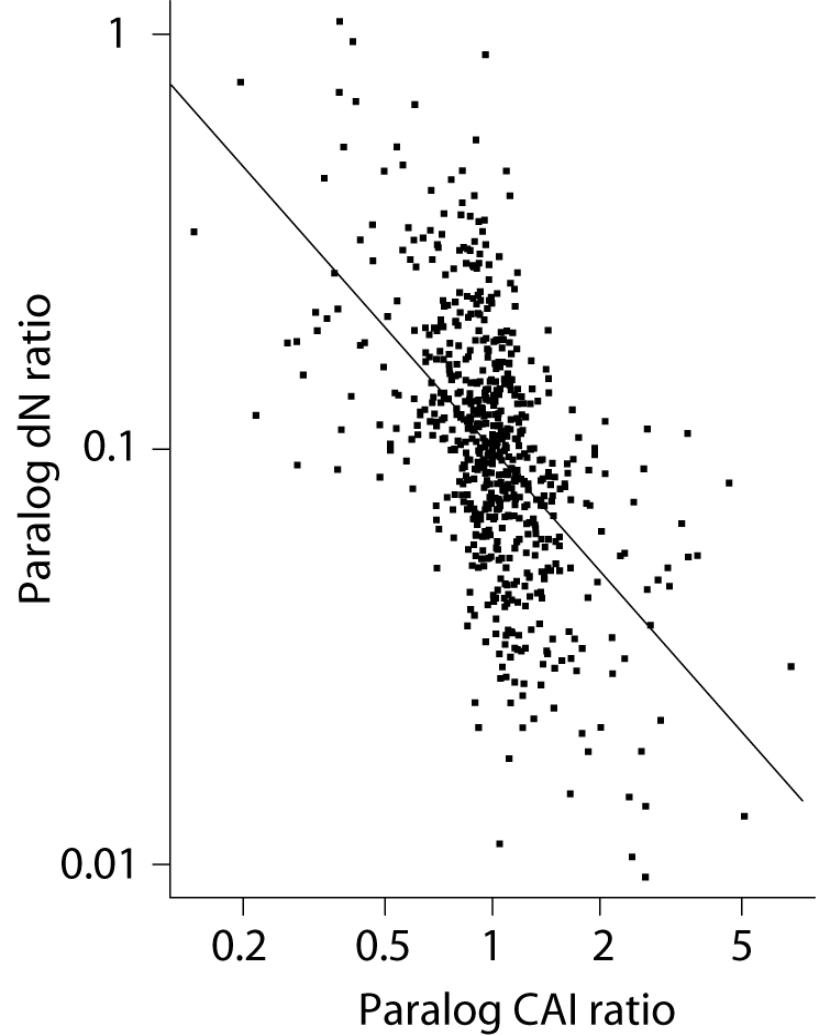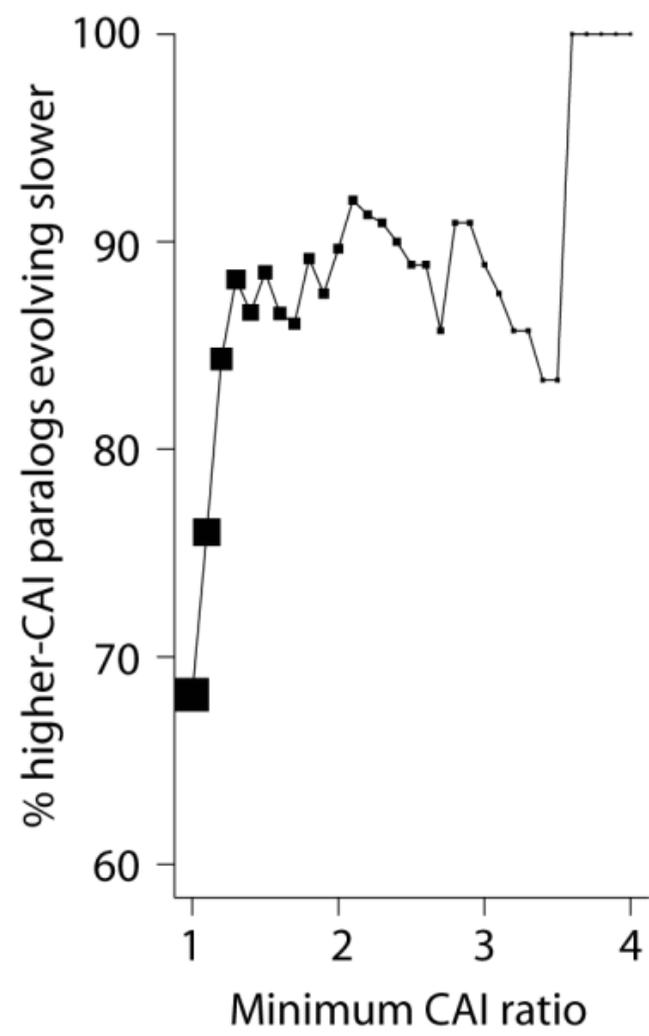